\documentclass[preprint,aps]{revtex4}

\usepackage{amsmath,amssymb,amsfonts,dcolumn,color,graphicx,graphics,setspace,latexsym,setspace,lscape,subfigure,placeins,epsfig,hyperref}

\newcommand{\be}{\begin{equation}}
\newcommand{\ee}{\end{equation}}
\newcommand{\ba}{\begin{eqnarray}}
\newcommand{\ea}{\end{eqnarray}}

\newcommand{\4}{\frac}

\newcommand{\lb}{\left}
\newcommand{\rb}{\right}
\providecommand{\norm}[1]{\lVert#1\rVert}

\begin{document}

\title{\Large \bf Bach flows of product manifolds}

\author{Sanjit Das and Sayan Kar}
\email{sanjit@cts.iitkgp.ernet.in, sayan@iitkgp.ac.in}
\affiliation{\rm Department of Physics and Meteorology {\it and} Centre for Theoretical Studies \\Indian Institute of Technology, Kharagpur, 721302, India}

\begin{abstract}
We investigate various aspects of a geometric flow defined using
the Bach tensor. 
Firstly, using a well-known split of the Bach tensor components for $(2,2)$
unwarped product manifolds, we solve the Bach flow equations for typical
examples of product manifolds like $S^2\times S^2$, $R^2\times S^2$. 
In addition, we obtain the fixed point condition for general 
$(2,2)$ manifolds and solve it for a restricted case.
Next, we consider warped manifolds.
For Bach flows on a special class of asymmetrically warped four manifolds,
we reduce the flow equations to a first order dynamical system, which
is solved exactly to find the flow characteristics. 
We compare our results for Bach flow with 
those for Ricci flow and discuss the differences 
qualitatively.
Finally, we conclude by mentioning possible directions for future work.   
\end{abstract}


\maketitle

\section{Introduction and overview} 
The Bach tensor \cite{bach,berg} is a traceless, symmetric, conformally invariant 
second rank tensor in
four dimensions which has the property of being divergence free. 
It is defined as:
\begin{equation}
B_{ik} = \nabla^j\nabla^l C_{ijkl} +\frac{1}{2} R^{jl}C_{ijkl}\label{bachdef}
\end{equation}
Using the contracted second Bianchi identity ( $\nabla^{l}R_{ijkl}=-\nabla_{i}R_{jk}+\nabla_{j}R_{ik}$ )
and contracted Hessian of the Ricci tensor ($\nabla^{l}\nabla_{i}R_{lk}=\nabla_{i}\nabla^{l}R_{lk}+R_{il}{R^{l}}_{k}-R_{ijkl}R^{jl}$) we can write Eqn.(\ref{bachdef}) as
\be
B_{ik}= \Box \lb( R_{ik} -\4{1}{6}R g_{ik} \rb)-\4{1}{3}\nabla_{i}\nabla_{k}R+\lb( C_{ijkl} + R_{ijkl} -R_{ij}g_{kl}\rb)R^{jl}
\ee
where $C_{ijkl}$ is the Weyl tensor and $R_{ik}$ is the Ricci tensor. 
We  mention that the above definitions are written using the 
Landau--Lifshitz sign 
convention \cite{ll} for the Riemann tensor (see \cite{conv} for a note on 
conventions and definitions).  
It is clear from the above definition, that the Bach tensor involves
four spatial derivatives as well as squares of second derivatives
of the metric tensor. It is, therefore a
higher derivative and higher order geometric object. 

In a recent article \cite{hlg}, it has been proposed that one
may define a geometric flow (of metrics on a manifold) using the
Bach tensor. The authors in \cite{hlg} suggest a flow equation,
following the equation for un--normalised Ricci flows \cite{ricci, hamilton, friedan,perelman}, given by
\begin{equation}
\frac{\partial g_{ij}}{\partial t} = \mp \frac{\kappa^2}{2} B_{ij}\label{bacheqn}
\end{equation}
where $t$ denotes the flow parameter (not the physical time). 

The above flow can be  shown as a gradient flow for the 
functional $\int_\mathcal{M}\norm{C_{ijkl}}^2\text{dm}$, where $\text{dm}$ is a fixed measure, on the manifold $\mathcal{M}/\text{Diff}\mathcal{M}$. 
This is similar to Ricci flow which is a gradient flow for the 
functional $\int_\mathcal{M}R~\text{dm}$\cite{glik}. The traceless property of the R. H. S. in the flow equation with the Bach tensor
makes the above flow different from unnormalised Ricci flow--in fact there 
is no distinction between normalised and un--normalised flows defined 
using the Bach tensor. 
The short-time existence for 
higher order geometric flows (including Bach flows) have been proved
very recently in \cite{bahu}.  
In our work here, we intend to understand the Bach flow of metrics through
explicit, illustrative examples. 

From a physics perspective, the Bach tensor arises in the well--known 
theory of conformal gravity (where instead of Einstein's equation one 
arrives at the Bach equation $B_{\mu\nu}=\kappa T_{\mu\nu}$ by varying the
Weyl--squared action w.r.t the metric tensor). Conformal gravity has been 
extensively studied 
--for details see \cite{mann1,mann2}. On the other hand, solutions of 
the Bach flow equation also
appear as solutions in the recently proposed 
five-dimensional  Horava--Lifshitz theory 
of gravity. 
Let us quickly recall the discussion on this aspect as given in \cite{hlg}. 

Consider five ($4+1$) dimensional Horava--Lifshitz gravity. Assume a
line element of the form:
\begin{equation}
ds^2 = - N^2 (t) dt^2 + g_{ij} \left (dx^i + N^i dt \right ) \left (dx^j + N^j dt\right )
\end{equation}
where $N (t)$ and $N^i$ are the lapse and shift functions, as assumed 
in a standard ADM split of the line element. The indices $i$, $j$ run from 
$1$ to $4$, since we are considering five spacetime dimensions.
The action for Horava--Lifshitz gravity with anisotropic scaling $z=D=4$
is given as
\begin{equation}
S_{HL} = \frac{2}{\kappa^2} \int dt d^4 x \sqrt{g} N K_{ij}G^{ijkl} K_{kl}
-\frac{\kappa^2}{8} \int dt d^4 x \sqrt{g} N \left (\frac{1}{\sqrt{g}} \frac{
\delta W}{\delta g_{ij}}\right ) G_{ijkl} \left (\frac{1}{\sqrt{g}} \frac{\delta
W}{\delta g_{kl}}\right ) 
\end{equation}
where $K_{ij}$ is the extrinsic curvature of the four dimensional
Riemannian manifold with metric
$g_{ij}$ and $G_{ijkl}$ is the de Witt metric in a superspace with parameter
$\lambda$. We take $\lambda <\frac{1}{4}$ so that the Euclidean action is 
bounded below. $W[g]$ is specified using detailed balance and is given as
\begin{eqnarray}
W[g] = W_{weyl} + W_{R^2} + W_{R} + W_{\Lambda_W} \\
=  \int d^4 x \left (a C_{ijkl} C^{ijkl} + b R^2  - cR - 2c\Lambda_W \right )
\end{eqnarray}
Solutions to $4+1$ Horava--Lifshitz gravity can thus be obtained from the 
geometric flow equation,
\begin{equation}
\frac{1}{N(t)} \frac{\partial g_{ij}}{\partial t} = \pm \frac{\kappa^2}{
2\sqrt{g}} G_{ijkl} \frac{\delta W[g]}{\delta g_{kl}} +\nabla_i \xi_j
+\nabla_j \xi_i
\end{equation}
Choosing $W[g]=W_{Weyl}$ and normalising the lapse to 1, one gets the
Bach flow equation quoted earlier. Thus, we note that the Bach flow
has a physically relevant origin and is obtainable via an
action principle.

The initial analysis in this article is based on a well-known result on the 
Bach tensor for the so-called $(2,2)$ product manifolds. The line element on
such manifolds are given as:
\begin{equation}
ds^2 = g_{\mu\nu}^{(1)}(x^\alpha)dx^\mu dx^\nu +  
g_{ab}^{(2)}(x^c)dx^a dx^b 
\end{equation}
where $x^{\mu}$ ($\mu=1,2$) and $x^a$ ($a=3,4$) are the coordinates
on  manifolds of dimension two. $g_{\mu\nu}$ and $g_{ab}$ depend
only on $x^{\mu}$ and $x^{a}$, respectively.
Such a construction is known as an unwarped product manifold--
the unwarped property
following from the fact that the metric tensor on the individual, 
two dimensional manifolds are dependent on the coordinates on each,
respective manifold, i.e. $g_{\mu\nu}^{(1)}(x^\alpha)$ does not depend
on $x^c$ and $g_{ab}^{(2)}(x^c)$ does not depend on $x^\alpha$.
For such line elements one can check \cite{fiedler,yano} that
the Bach tensor splits as follows:
\begin{eqnarray}
B_{\mu\nu} = \frac{1}{3} \nabla_\mu\nabla_\nu {}^{(1)}R -\frac{1}{3} g_{\mu\nu}^{(1)} \left \{ \nabla^\alpha \nabla_\alpha {}^{(1)}R -\frac{1}{2} \nabla^a\nabla_a {}^{(2)}R +\frac{1}{4}\left ({}^{(1)}R^2 -{}^{(2)}R^2\right )\right \} \\
B_{ab} = \frac{1}{3} \nabla_a\nabla_b {}^{(2)}R -\frac{1}{3} g_{ab}^{(2)} \left \{ \nabla^a \nabla_a {}^{(2)}R -\frac{1}{2} \nabla^\mu\nabla_\mu {}^{(1)}R +
\frac{1}{4}\left ({}^{(2)}R^2 -{}^{(1)}R^2\right )\right \} 
\end{eqnarray}
Hence, the Bach tensor is exclusively determined in terms of the
derivatives of the Ricci scalars of the individual two dimensional
manifolds in the product.

Our article is organised as follows. In Section II
we analyse Bach flows on simple, unwarped product manifolds using the
above mentioned decomposition. Further, in Section III, 
we obtain the fixed point equation
of Bach flows on unwarped product manifolds and find some analytical, as well 
as numerical solutions.  
In Section IV, we investigate what might
happen if we considered warped spacetimes, in paricular four dimensional
Lorentzian spacetimes which are asymmetrically warped. It turns out
that for a specific family of warped spacetimes, the flow equations
reduce to a dynamical system which can be analytically solved. 
Finally, in Section V, we conclude with a summary and some
perspectives.
   
\section{Bach flows on $S^2\times S^2$ and $S^2\times R^2$}

The simplest example of unwarped (2,2) product manifolds are 
$S^2\times S^2$ and $S^2\times R^2$. 
Let us discuss the $S^2\times S^2$ case first.

In order to discuss geometric flows, we assume that the line element
on the $S^2\times S^2$ product manifold is given as
\begin{equation}
ds^2 = A^2 (t) ds_{(1)}^2 + B^2 (t) ds_{(2)}^2
\end{equation}
where $A^2(t)$ and $B^2(t)$ are the scale factors, dependent on the
flow parameter $t$, and $ds^2_{(1,2)}$ are the line elements on the
round two-sphere of unit radius.

The Bach flow equations, using the split mentioned in the Introduction,
turn out to be
\begin{eqnarray}
\frac{df}{dt} = \pm \frac{\kappa^2}{24} \left (\frac{4}{f^2}-\frac{4}{g^2}\right ) f \\
\frac{dg}{dt} = \pm \frac{\kappa^2}{24} \left (\frac{4}{g^2}-\frac{4}{f^2}\right ) g 
\end{eqnarray}
where $f(t) = A^2(t)$ and $g(t)=B^2(t)$.

It is easy to note (since the Bach tensor is traceless) that
\begin{equation}
f(t) g(t) =  A^2(t) B^2(t) = Constant
\end{equation}
i.e. $A^2 B^2$ is conserved along the flow.
This aspect clearly distinguishes the Bach flow from un--normalised Ricci flows.
Secondly, the solutions for $f$ and $g$ (using the $+$ sign in the
above equations) are obtained as:
\begin{eqnarray}
f^2(t) = A^4(t) = C_1 \tanh \left (\frac{2t'}{C_1} + C_2 \right ) \\
g^2(t) = B^4(t) = C_1 \coth \left (\frac{2t'}{C_1} + C_2 \right ) 
\end{eqnarray}
where $C_1$ and $C_2$ are constants and $t'$ is the
$t$ rescaled with $t'=\frac{\kappa^2}{6}t$.

Thus, $A(t)$ has a past singularity (curvature diverges) 
but exists eternally in the future.
On the other hand $B(t)$ is infinitely large at the $t$ value where
$A(t)$ becomes zero, but, asymptotically, goes to the same constant
as $A(t)$. The product $A^4(t) B^4(t)= C_1^2$.
\begin{figure}
	\centering
	\subfigure[$A_{0}=2$ , $B_{0}=5$ for $S^{2}\times S^{2}$]{\includegraphics[width=0.6\textwidth]{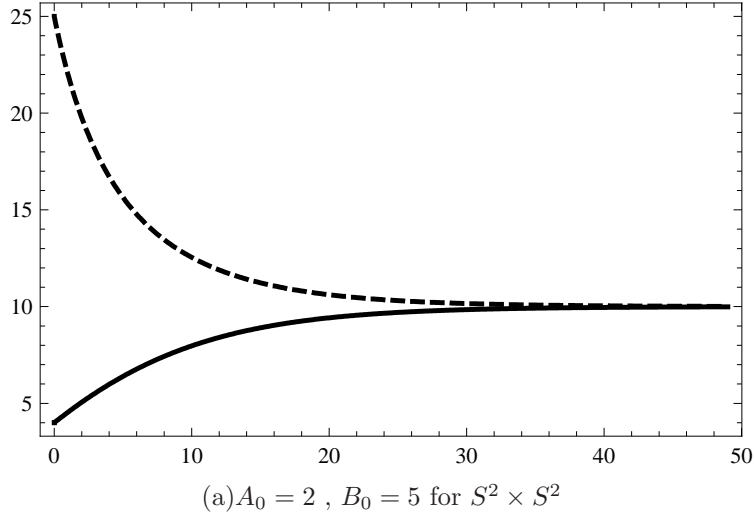}\label{figs2xs2}}
	
	\caption{$A^{2}(t)\text{(continuous)}$ and $B^{2}(t)\text{(dashed)}$.The horizontal axis is $t$, $T_{s}= -6.35$.}
	\label{s2xs2}
	\end{figure}

We have numerically solved the Bach flow equations for some particular 
initial values of $A_{0}$, $B_{0}$ as shown in Fig.(\ref{s2xs2}). 
Note that the generic features are quite different from Ricci flows
of similar product manifolds, which we have discussed in one of
our earlier works \cite{sanjit}.

\begin{figure}
	\centering
	\subfigure[$A_{0}=5$ , $B_{0}=2$, $T_s=-37.5$ for $S^{2}\times R^{2}$]{\includegraphics[width=0.6\textwidth]{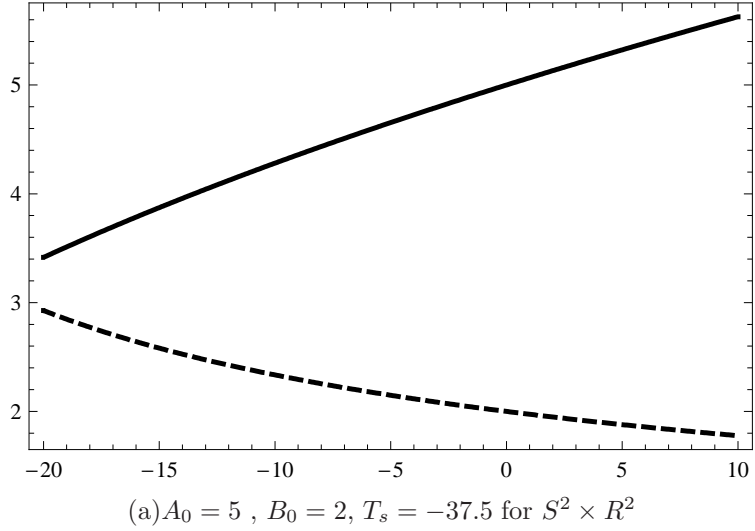}\label{figs2xr2}}
	
	\caption{$A^{2}(t)\text{(continuous)}$ and $B^{2}(t)\text{(dashed)}$.The horizontal axis is $t$.}
	\label{s2xr2}
	\end{figure}

Let us now turn to $S^2\times R^2$. Here one of the Ricci scalars is
zero, and the flow equations are even more simple.  We have
\begin{eqnarray}
\frac{df}{dt} =\pm \frac{1}{3} \frac{1}{f} \\
\frac{dg}{dt} =\mp \frac{1}{3} \frac{g}{f^2} 
\end{eqnarray}
The solutions for $f$ and $g$ are given as,
\begin{eqnarray}
f(t) = A^2(t) = \sqrt{t \mp \frac{3C}{2\kappa^2}}
\\
g(t) = B^2(t) = \frac{1}{\sqrt{t \mp \frac{3C}{2\kappa^2}}}
\end{eqnarray}
where $C$ is a constant of integration and as before, $A^2 B^2$ is conserved
along the flow.
Here too we note that while $A^2(t)$ is infinite at some $t$, $B^2(t)$
becomes zero there. Fig.(\ref{s2xr2}) shows the numerical evolution of 
the scale factors for some particular initial values. Once again the 
results are very different from the standard Ricci flow evolution \cite{sanjit} 

We may also consider products like $S^2\times H^2$, $ H^2\times H^2$
and $R^2\times H^2$. However, there is no change in the results for
the above-mentioned manifolds with constant curvature metrics on them.
This is because the Bach tensor depends on the square of the
Ricci scalar of the individual two manifolds--thus having negative
curvature does not yield anything new. In other words, Bach flow
on (2,2) manifolds of constant curvature do not {\em see} the
sign of the Ricci scalar, of the individual two manifolds. 
This is another major point of difference with Ricci flows
where differences arise if one considers products like $H^2\times H^2$
or $H^2\times R^2$ or $H^2\times S^2$ (see \cite{sanjit} for details).
It is also worth noting that on an Einstein manifold the Bach tensor is zero. 
So for constant curvature four manifolds like $S^{4}$ and $H^{4}$ there 
is no Bach flow. 
\section{Fixed points in Bach flow}
In this section we will find the fixed points of Bach flows on 
$\mathcal{M}^2\times \mathcal{M}^2$ manifolds. The fixed point
equations are trivial and follow from 
the equations,
\begin{equation}
B_{\mu\nu}=0 \hspace{0.2in};\hspace{0.2in} B_{ab}=0
\end{equation}
Taking the trace of the above equations w.r.t $g^{\mu\nu}$ and $g^{ab}$ and 
after some straightforward algebra, we find that the
conditions for fixed points turn out to be
\begin{eqnarray}
\nabla_\mu\nabla_\nu {}^{(1)}R -\frac{1}{2} g_{\mu\nu} \nabla^\alpha\nabla_\alpha {}^{(1)}R = 0\\
\nabla_a\nabla_b {}^{(2)}R -\frac{1}{2} g_{ab} \nabla^c\nabla_c {}^{(2)}R = 0
\end{eqnarray}
It may be mentioned here that, each of the above conditions  
can be recast as a conformal Killing equation for the conformal
Killing vector $\nabla_\nu {}^{(1)}R$ or $\nabla_a {}^{(2)}R$ \cite{date}.
Thus, the Ricci scalars of the individual two dimensional manifolds must
necessarily satisfy a constraint, in order to be fixed points. We will
now explore these constraints in greater detail.

Let us assume the generic metric on the $(2,2)$ product manifold as
\begin{equation}
ds^2 = \Omega_1^2 (x,y, t) \left (dx^2+ dy^2 \right ) +
\Omega_2^2 (x', y',t) \left (dx'^2+dy'^2\right )
\end{equation}
i.e. each two--manifold, in the chosen coordinates, is conformally flat. 
Using the fixed point condition for either one of them, we obtain
\begin{equation}
\nabla_{\mu}\nabla_{\nu} \left [ e^{-\eta_1} \nabla_E^2 \eta_1 \right ] -
\frac{1}{2} \Omega_1^2 \delta_{\mu \nu} \nabla^{\alpha} \nabla_{\alpha} \left [ e^{-\eta_1}
\nabla_E^2 \eta_1 \right ] = 0
\end{equation}
where $\eta_1=\ln \Omega_1^2$ and $\nabla_E^2$ is the Laplacian
in Euclidean space ($\Omega_{(1,2)}^2 =1$).
A similar equation exists for $\Omega_2^2$ as well.

From the fixed point condition one can also arrive at a pair of partial
differential equations involving the Ricci scalar and the conformal factor. 
These are
\begin{eqnarray}
\frac{\partial^2 {}^{(1)}R}{\partial x^2} -
\frac{\partial^2 {}^{(1)}R}{\partial y^2} -
\frac{\partial\eta_1}{\partial x}
\frac{\partial {}^{(1)}R}{\partial x} +
\frac{\partial\eta_1}{\partial y}
\frac{\partial {}^{(1)}R}{\partial y} =0 \\
\frac{\partial^2 {}^{(1)}R}{\partial x\partial y} -
\frac{1}{2}\frac{\partial\eta_1}{\partial y}
\frac{\partial {}^{(1)}R}{\partial x} - 
\frac{1}{2}\frac{\partial\eta_1}{\partial y}
\frac{\partial {}^{(1)}R}{\partial x} =0 
\end{eqnarray}
Similar equations also exist for $\Omega_2^2$ (or $\eta_2$ and
${}^{(2)}R$). 

To find a simple solution, one can assume $\eta_1$ (or $\Omega_1$) 
to be a function of $x$ alone. In that case, the second equation
is automatically satisfied. The first equation, on the other hand,
gives,
\begin{equation}
\frac{d^2 {}^{(1)}R}{dx^2} -\frac{d\eta_1}{dx} \frac{d {}^{(1)}R}{dx}=0
\end{equation}
Assuming
\begin{equation}
\alpha = \frac{d{}^{(1)}R}{dx}
\end{equation}
we get,
\begin{equation}
\alpha = C_1\Omega_1^2
\end{equation}
Substituting 
\begin{equation}
{}^{(1)}R = -e^{-\eta_1}\frac{d^2\eta_{1}}{dx^2}\label{scalarformula}
\end{equation}
we obtain the following equation for $\eta_1$.
\begin{equation}
\eta_1'''-\eta_1'' \eta_1'= C_2 e^{2\eta_{1}}
\end{equation}
where the prime denotes differentiation w.r.t. $x$ and $C_2=-C_1$.
We can find a solution with the following ansatz,
\begin{equation}
\eta_1 = \ln \Omega_1^2=\beta \ln x
\end{equation}
Substituting this back in the third--order differential equation, we obtain
\begin{equation}
\frac{2\beta}{x^3}+\frac{\beta^2}{x^3} = C_2 x^{2\beta}
\end{equation}
One can check that we have a consistent solution if
\begin{equation}
\beta=-\frac{3}{2} \hspace{0.2in};\hspace{0.2in} C_2=2\beta+\beta^2=-\frac{3}{4}\end{equation}
Therefore, we have,
\begin{equation}
\Omega_1^2 = x^{\beta} = x^{-\frac{3}{2}} 
\end{equation}
Similarly, assuming $\Omega_2^2$ as a function of $x'$ alone we will
end up with the same solution for $\Omega_2^2$ with $x$ replaced by $x'$.
The resulting line element is given as
\begin{equation}
ds^2 = \frac{1}{x \sqrt x} \left (dx^2 + dy^2\right ) + \frac{1}{x'\sqrt{x'}}
\left (dx'^2 +dy'^2\right )
\end{equation}
The Ricci scalar, R, of the four manifold, turns out to be
\begin{equation}
R = -\frac{3}{2} \left ( \frac{1}{\sqrt{x}} +\frac{1}{\sqrt{x'}} \right )
\end{equation}
  
\begin{figure}
	\centering
	\subfigure[$\eta(t)$ (continuous), $\xi(t)$ (dashed) \text{and} $\zeta(t)$ (dotted) vs $t$]{\includegraphics[width=0.5\textwidth]{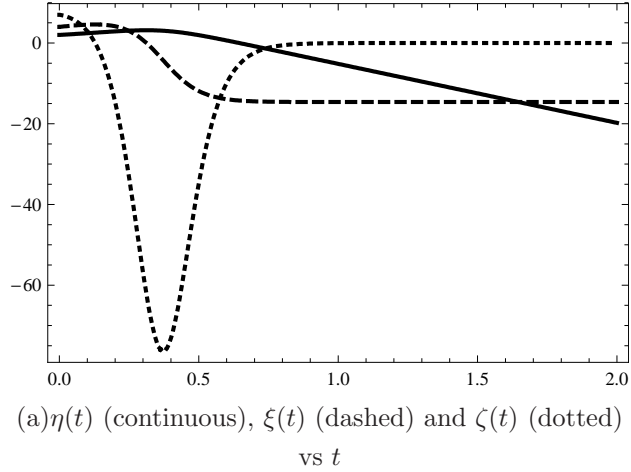}\label{figfixed1_1}}
	\subfigure[The scalar curvature ${}^{(1)}R$ vs. t]{\includegraphics[width=0.5\textwidth]{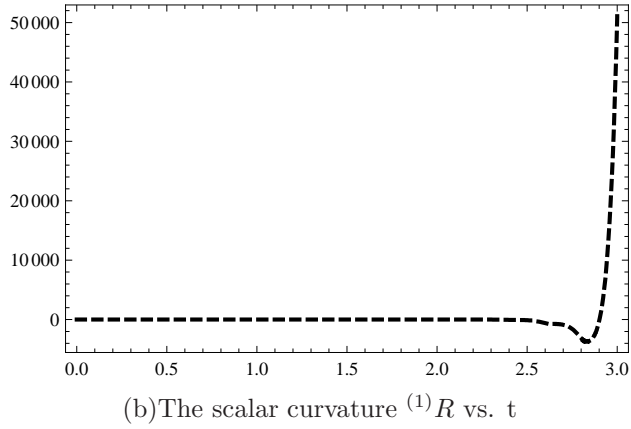}\label{figfixed1_2}}
	
	\caption{Fixed point analysis on $\mathcal{M}^{2}\times \mathcal{M}^{2}$for the restricted case mentioned in the text.}
	\label{fixed1}
	\end{figure}

One can further convert the third order equation into a dynamical system
by defining $\eta_1'=\eta'=\xi$ (we have renamed $\eta_1$ as $\eta$). We have
\begin{eqnarray}
\eta' = \xi \\
\xi'=\zeta \\
\zeta'= \zeta\xi+ C_2 e^{2\eta}
\end{eqnarray}
This system can now be solved numerically for the general case
to obtain desired results. 
We have taken initial values as $\eta_{0}=2$, $\xi_{0}=5$, $\zeta_{0}=7$, 
$C_2=-0.75$
and numerically solved the dynamical system. The results are 
shown in Fig.(\ref{figfixed1_1}). Additionally, the  variation of the 
scalar curvature ${}^{(1)}R =-e^{-\eta}\zeta$ is also found 
from Eqn.(\ref{scalarformula}) and 
is shown in Fig.(\ref{figfixed1_2}). ${}^{(1)}R$ diverges to positive infinity
because $\zeta$ becomes zero (from the negative side) for large $t$ and
$e^{-\eta}$ grows to very large positive values ($\eta$ tends to large
negative values for large $t$).  

\section{Asymmetrically warped products}

Let us now turn to a discussion on warped products. As a prototype
family of warped product four manifolds let us assume the line element as
\begin{equation}
ds^2=-A^2(t) e^{2 f(\sigma)} d\tau^2 + B^2 (t) e^{2 g(\sigma)} \left [
dx^2 +dy^2\right ] + C^2 (t) d\sigma^2
\end{equation}
Topologically the above manifold is $\mathbb R^3\times \mathbb R$ or $\mathbb R^3\times \mathbb S^1$
or $\mathbb R^3\times \mathbb S^1/ \mathbb Z_2$ depending on the domain of the coordinate $\sigma$.
The warping introduces curvature properties.
The above line element is largely motivated by the so--called brane--world
models with asymmetric warping \cite{brane}, though with one dimension less. 
Note that $f\neq g$ is necessary in order to have a non--zero Bach tensor.
To keep things simple let us assume further
\begin{equation}
f (\sigma) = k_1 \sigma \hspace{0.2in};\hspace{0.2in} g(\sigma) = k_2 \sigma
\end{equation}
We now find the Bach tensor and, subsequently the Bach flow equations
for the $A^2$, $B^2$ and $C^2$.

The Bach tensor components for the abovementioned line element are,
\ba
B_{tt}~=~-\4{A^2}{3C^4}\alpha e^{2k_{1}\sigma}\\
B_{xx}~=B_{yy}=~-\4{B^2}{3C^4}\beta e^{2k_{2}\sigma}\\
B_{zz}~=~\4{1}{3C^2}\gamma 
\ea 
where $\alpha, \beta$ and $\gamma$ are given as,  
\ba
\alpha~=~k_{1}\lb(k_{1}-k_{2}\rb)\lb(k_{1}-4k_{2} \rb)\lb(k_{1}+3k_{2} \rb)\\
\beta~=~k_{1}\lb(k_{1}-k_{2}\rb)\lb(k_{1}-4k_{2} \rb)\lb(k_{1}+k_{2} \rb)\\
\gamma~=~k_{1}\lb(k_{1}-k_{2}\rb)\lb(k_{1}-4k_{2} \rb)\lb(k_{1}-k_{2} \rb)
\ea
It is easy to check that the above components satisfy the traceless
and divergence free condition on the Bach tensor. Tracelessness of Bach 
tensor gives contraint on $\alpha$, $\beta$ and $\gamma$ as 
\be
\alpha - 2\beta + \gamma =0\label{constr}
\ee
We can further verify that it is  divergence free($\nabla^{i}B_{ik}=0$). All the components of $\nabla^{i}B_{ik}$ trivially go to zero expect for $k=3$. 
Therefore, we have only one non--trivial expression given as 
\be
\nabla^{i}B_{i \sigma}=-\4{2}{3C^{4}}\lb[ k_{1}(\alpha - \beta) + k_{2} (\alpha - 3 \beta) \rb]
\ee
which is essentially zero after substituting the values of $\alpha$ and $\beta$.

We also note that the Bach tensor is zero when $k_1=k_2$ and when $k_1=0$.
The former is a trivial conformally flat case. For the latter, i.e. with
$k_1=0$, conformal flatness is not directly visible. However, we can see
it as follows (assume that $A$, $B$ and $C$ are scaled into the coordinates).
\begin{eqnarray}
ds^2 = -dt^2 + e^{2k_2\sigma} \left (dx^2 + dy^2\right ) + d\sigma^2
\\
=e^{2k_2\sigma} \left [ -e^{-2k_2\sigma} dt^2 + dx^2+dy^2 + e^{-2k_2\sigma}d\sigma^2 \right ]\\
=\frac{1}{k_2\sigma'^2} \left [ \sigma'^2 dt'^2 + d\sigma'^2 + dx^2+dy^2 \right ]\\
=\frac{1}{\eta^2-\xi^2} \left [ -d\eta^2+d\xi^2 +d\sigma'^2 + dx^2+dy^2 \right ]
\end{eqnarray}
where $t' =k_2 t$ and, in the last step we have used a two dimensional Rindler
like transformation from $\sigma',t'$ to $\eta,\xi$. It is interesting to note that for $k_{1}=4k_{2}$ all the components of Bach tensor are zero, i.e the metric is Bach flat. However, it is not conformally flat for  $k_{1}=4k_{2}$ because of the presence of non vanishing components of Weyl tensor. 
Thus, as is well--known, for metrics on manifolds of dimension $n\ge 4$ 
the vanishing of Weyl tensor components ensures that the metric
is conformally flat (and Bach flat, for $n=4$) but a zero Bach tensor does not
necessarily imply conformal flatness in four dimensions.

Using the above, we can easily write down the Bach flow equations as,
\ba
\4{dA^2}{dt}~=~\mp \4{\kappa^{2}}{2}\alpha \4{A^2}{3C^4}\\
\4{dB^2}{dt}~=~\pm \4{\kappa^{2}}{2}\beta \4{B^2}{3C^4}\\
\4{dC^2}{dt}~=~\mp \4{\kappa^{2}}{2}\gamma \4{1}{3C^2}\\
\ea

We can absorb  $\4{\kappa^2}{6}$ by rescaling  $t$ and easily obtain 
the exact solutions for the dynamical system.
Here we have two sets of solutions depending on the sign used in the
R. H. S. of the flow equations. Thus, 
\ba
C^{2}(t)= \lb( 2C_{0}^{4}-2\gamma t \rb)^{1/2}~~~\text{for $t<\4{C_{0}^{4}}{\gamma}$} \\
B^{2}(t)= B_{0}^{2} \lb( 2C_{0}^{4}- 2 \gamma t \rb)^{-\beta/2\gamma} ~~~\text{for $t<\4{C_{0}^{4}}{\gamma}$} \\
A^{2}(t)=A_{0}^{2}}{\lb(2 C_{0}^{4}-2\gamma t  \rb)^{\alpha/2\gamma}~~~\text{for $t<\4{C_{0}^{4}}{\gamma}$} 
 \ea
For the backward flow, the solutions will be:
\ba
C^{2}(t)= {\lb(2 C_{0}^{4}+2\gamma t \rb)}^{1/2}\\
B^{2}(t)= B_{0}^{2}{\lb( 2C_{0}^{4}+2\gamma t \rb)}^{-\beta/2\gamma}\\
A^{2}(t)=A_{0}^{2}{\lb( 2C_{0}^{4}+2\gamma t  \rb)}^{\alpha/2\gamma}
\ea
The expressions of $\alpha$, $\beta$ and $\gamma$ reveal that they vanish 
for $k_1=0$,$k_1=k_2$or $k_1=4k_2$. However, for non-vanishing $\gamma$, 
$\alpha$ or $\beta$ can still be zero if $k_1=-3k_2$ or $k_1=-k_2$, 
respectively. 
These cases are quite easy to understand from the Bach flow equations. 
The solutions will be the same as quoted earlier except that 
either $A(t)=\text{const.}$ (for $\alpha=0$) or  
$B(t)=\text{const.}$ (for $\beta=0$).  
We can choose values of $k_1$ and $k_2$ to explore the solutions 
of the evolution equations explicitly. We have accounted for two 
different regimes of 
$k_1$ and $k_2$: $k_1>4k_2$ and $k_1<4k_2$ respectively for forward and 
backward 
flows. Even $k_1<4k_2$ has two sub--domains -- $k_1<k_2$ and $k_2<k_1<4k_2$.  
Below, we discuss each of the abovementioned cases in detail.
In the corresponding figures, $A^{2}(t)$, $B^{2}(t)$ and $C^{2}(t)$ are denoted by 
a bold line, a dashed line and a dotted line, respectively. 
The initial conditions 
are taken as $A_0^{2}=2$, $B_0^{2}=5$ and $C_0^{2}=7$. 
The left and right figures in  
Fig.\ref{warpfig} are for forward and backward flows respectively. 

\subsection{$\alpha, \beta, \gamma \ne 0$: $\bf k_1>4k_2$}
In Fig.\ref{warpfig1} we have taken $k_1=8$ and $k_2=1$ for which $(\alpha,\beta,\gamma)=(2464,2016,1568)$. In this regime of 
$k_1$ and $k_2$ both $A^2(t)$ and $C^{2}(t)$ are decreasing whereas 
$B^2(t)$ is increasing. Thus, it is an ancient solution and develops a 
future singularity at $T_s=0.016$. Contrarily, the backward flow emerges
as an immortal kind of flow with a past singularity at  $T_s=-0.016$. 
Here $B^2(t)$ is decreasing while $A^2(t)$ and $C^2(t)$ increase 
and 
eventually diverge after crossing each other (checked, but not shown in figure)
at some time in future. 
We also note that $B^2(t)$ and $A^2(t)$ cross
each other and diverge for larger $t$. 

\subsection{$\alpha, \beta, \gamma \ne 0$: ${\bf k_1<4k_2}(k_1<k_2)$}

Fig.\ref{warpfig2} shows a sub-regime 
of $k_1$ and $k_2$ where $k_1<k_2$. Here we choose $k_1=1$ and $k_2=5$ for which $(\alpha,\beta,\gamma)=(1216, 456, -304)$. In the forward flow, 
both $B^2(t)$ and $C^2(t)$ deviate 
from each other after meeting at a point. $A^2(t)$ decays and asymptotically 
reaches a constant value. This is an immortal flow having a past singularity 
at $T_s=-0.081$. On the other hand, the backward flow is an ancient solution 
having a future singularity at $T_s=0.081$.\\
\subsection{$\alpha,\beta,\gamma \ne 0$: ${\bf k_1<4k_2}(k_2<k_1<4k_2)$}

Fig.\ref{warpfig2_1} shows the other sub-regime 
of $k_1$ and $k_2$ where $k_2<k_1<4k_2$. Here we choose $k_1=7$ and $k_2=3$ for which $(\alpha,\beta,\gamma)=(-2240, -1400, -560)$. In the forward flow, 
both $A^2(t)$ and $C^2(t)$ deviate 
from each other after crossing at some $t$. $B^2(t)$  crosses $C^{2}(t)$ 
and asymptotically 
reaches a constant value. $A^{2}(t)$, while increasing, 
crosses both $B^{2}(t)$ and $C^{2}(t)$ 
at two different $t$ values. It is an immortal flow having a past singularity 
at $T_s=-0.0437$. On the other hand, the backward flow is an ancient solution 
having a future singularity at $T_s=0.0437$. \\

\subsection{$\alpha=0$ ,$\beta, \gamma \ne 0$: ${\bf k_1<4k_2}(k_1<k_2)$}

We will now focus on a special case where $\beta$, $\gamma$ are non zero but 
$\alpha$ is zero. The figure on the right in 
Fig.\ref{warpfigsp} depicts this case. This is possible for 
$k_1=-3k_2$ which lies in the regime of $k_1<k_2$. This is an ancient 
solution with future singularity at $T_s=0.073$. We can see that the
vanishing of $\alpha$ flips the nature of the solution from immortal to 
ancient. 

\subsection{$\beta=0$,$\alpha,\gamma \ne 0$:${\bf k_1<4k_2}(k_1<k_2)$}
A similar feature appears for $\beta =0$ (i.e $k_1=-k_2$) where the singularity is at $T_s=1.22$. 

A summary of 
the different regimes and the corresponding flow characteristics 
is shown in 
Table \ref{tablecompare}.

\begin{table}[htbp]
\begin{center}
\begin{tabular}{||c|c|c||}
   \hline
Regime                      & Forward                      &                    Backward \\
\hline\hline
$k_1>4k_2$      & ancient                       & immortal\\
$\alpha,\beta,\gamma\neq 0 $& &\\
\hline
$k_1<k_2$      & immortal                      & ancient\\
$\alpha,\beta,\gamma\neq 0 $& &\\
\hline
$k_2<k_1<4k_2$   & immortal               & ancient\\
$\alpha,\beta,\gamma\neq 0 $& &\\
\hline
$k_1<k_2$      & ancient                     & immortal\\
$\alpha= 0,\beta,\gamma\neq 0 $& &\\
\hline
$k_1<k_2$      & ancient                     & immortal\\
$\beta= 0,\alpha,\gamma\neq 0 $& &\\
\hline\hline

\end{tabular}
\end{center}
\caption{The various cases in Bach flow of asymmetrically warped product manifolds}
        \label{tablecompare}
\end{table}

\begin{figure}
	\centering
	\subfigure[$A^{2}(t)(continuous)$, $B^{2}(t)(dashed)$, $C^{2}(t)(dotted)$ v/s $t$ for $k_1>4k_2$]
{
\includegraphics[width=0.4\textwidth]{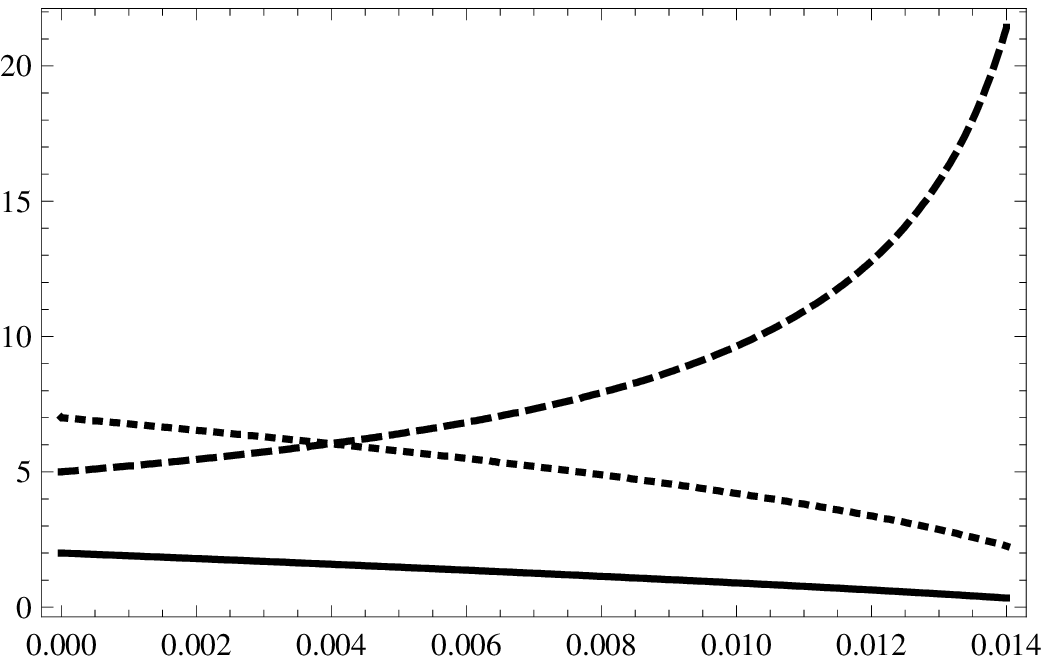} 
\includegraphics[width=0.4\textwidth]{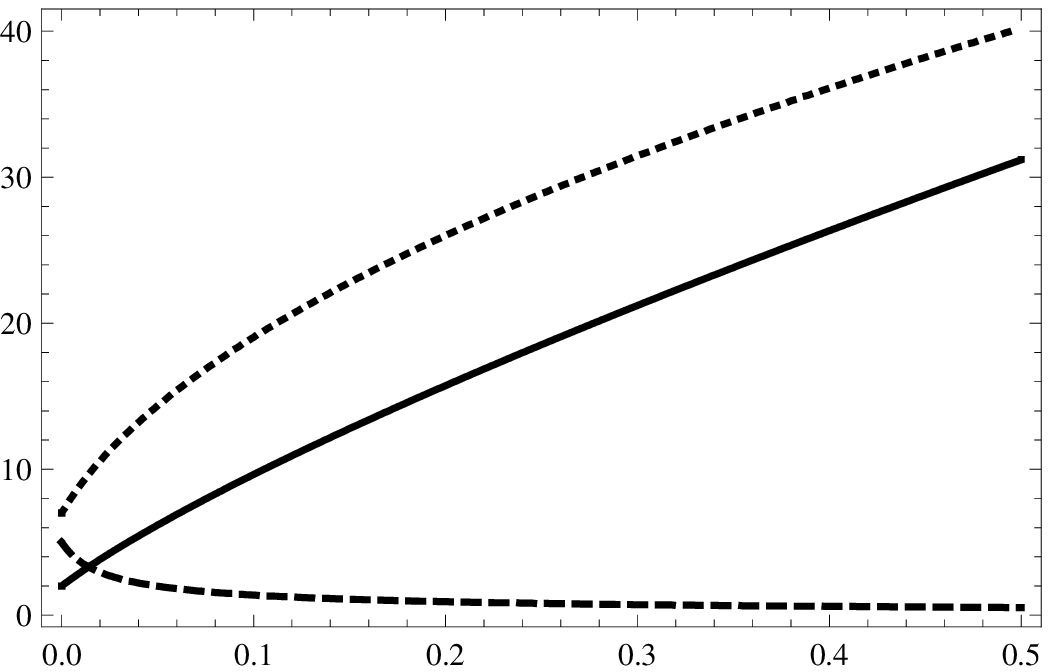} 
\label{warpfig1}
}
\subfigure[$A^{2}(t)(continuous)$, $B^{2}(t)(dashed)$, $C^{2}(t)(dotted)$ v/s $t$ for $k_1<k_2$]
{
\includegraphics[width=0.4\textwidth]{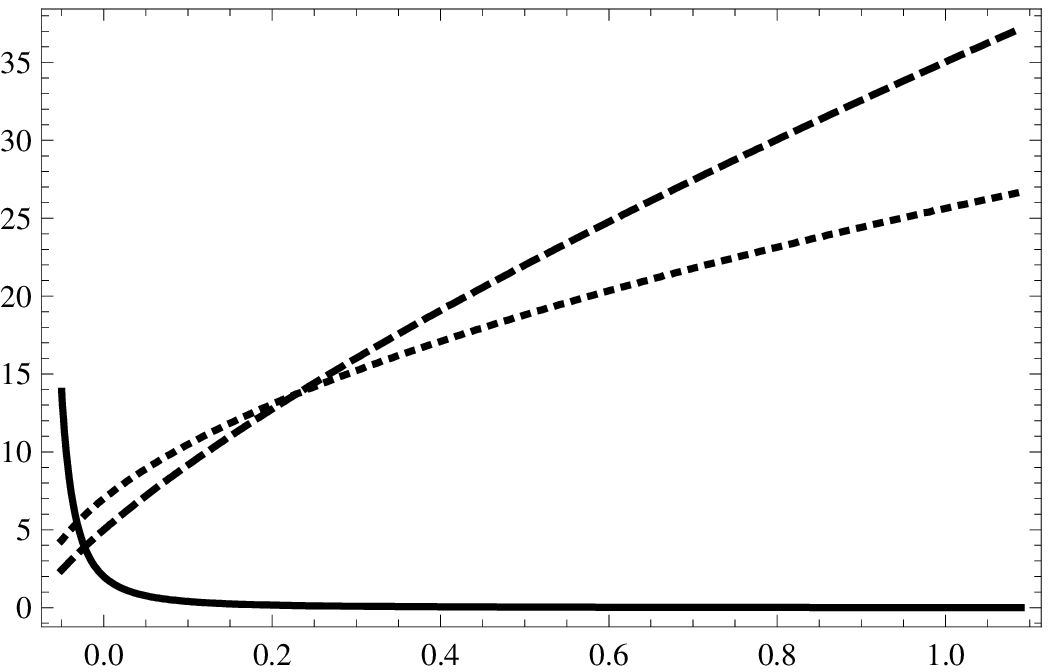} 
\includegraphics[width=0.4\textwidth]{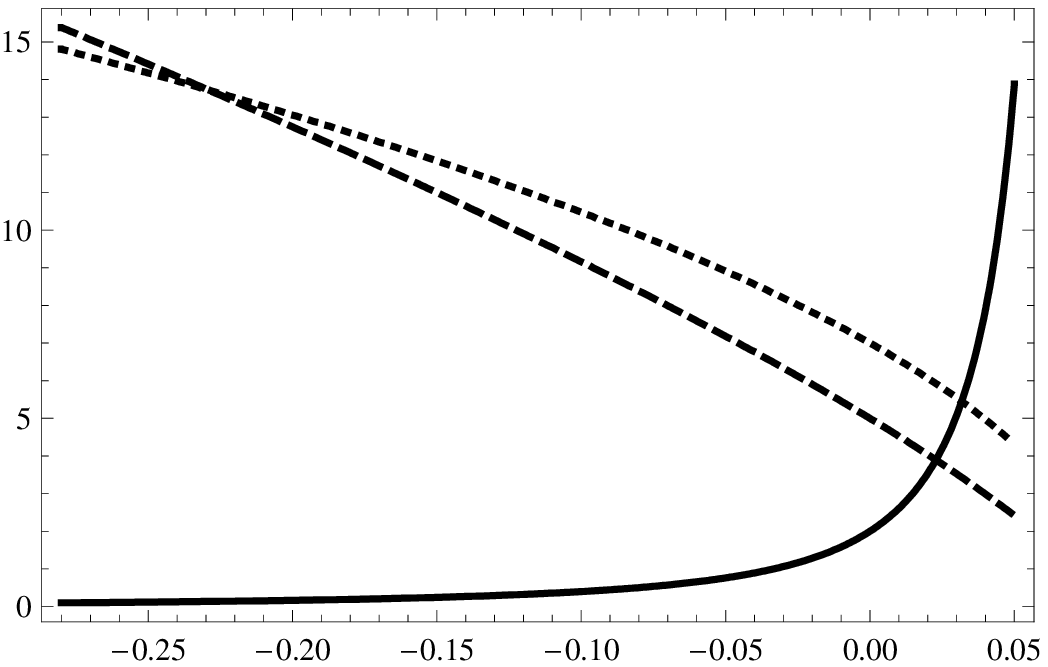}
\label{warpfig2}
}
\subfigure[$A^{2}(t)(continuous)$, $B^{2}(t)(dashed)$, $C^{2}(t)(dotted)$ v/s $t$ for $k_2<k_1<4k_2$]
{
\includegraphics[width=0.4\textwidth]{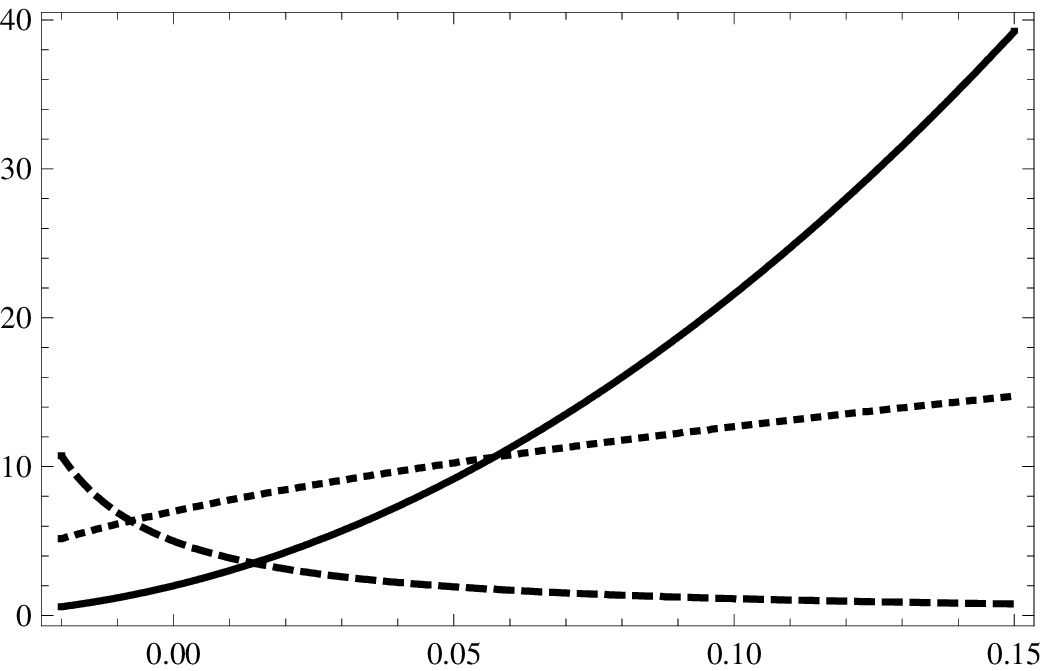} 
\includegraphics[width=0.4\textwidth]{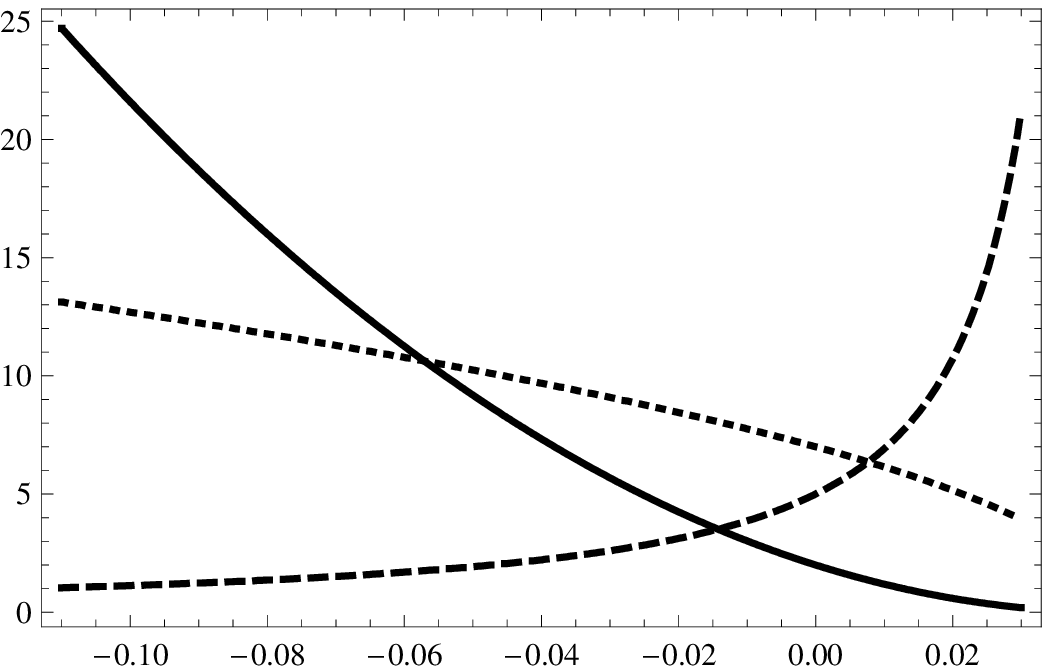}
\label{warpfig2_1}
}
	\caption{Bach flow : Forward and backward flow 
for different values of $k_1$ and $k_2$}
	\label{warpfig}
	\end{figure}
	
	\begin{figure}[h!]
	\centering
	\subfigure[$A^{2}(t)(continuous)$, $B^{2}(t)(dashed)$, $C^{2}(t)(dotted)$ v/s $t$  for $k_1<4k_2$]
{
\includegraphics[width=0.4\textwidth]{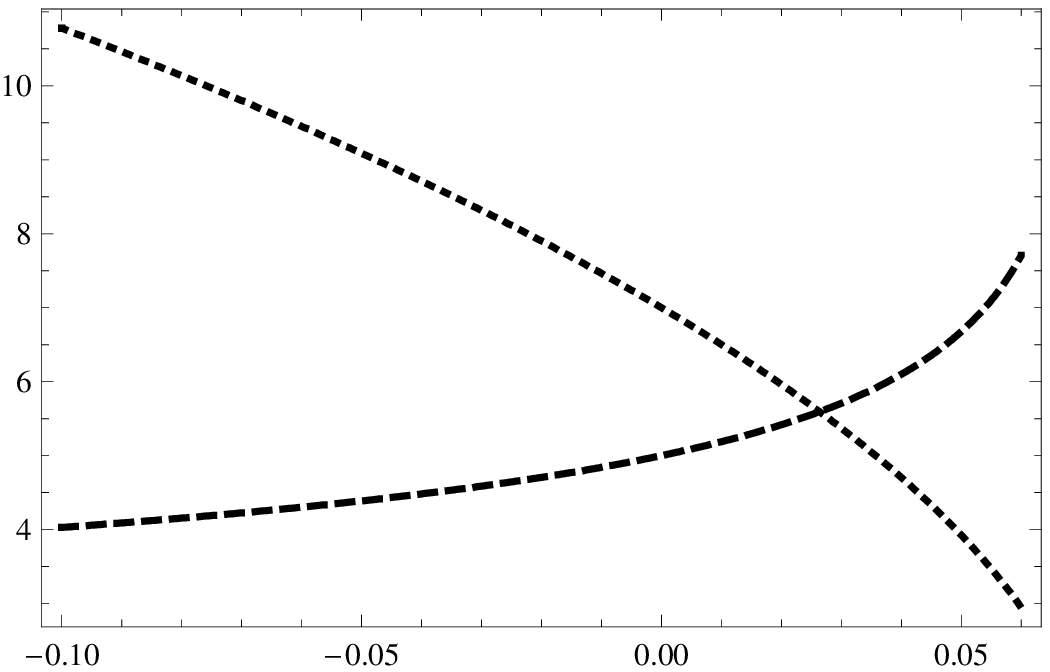} 
\includegraphics[width=0.4\textwidth]{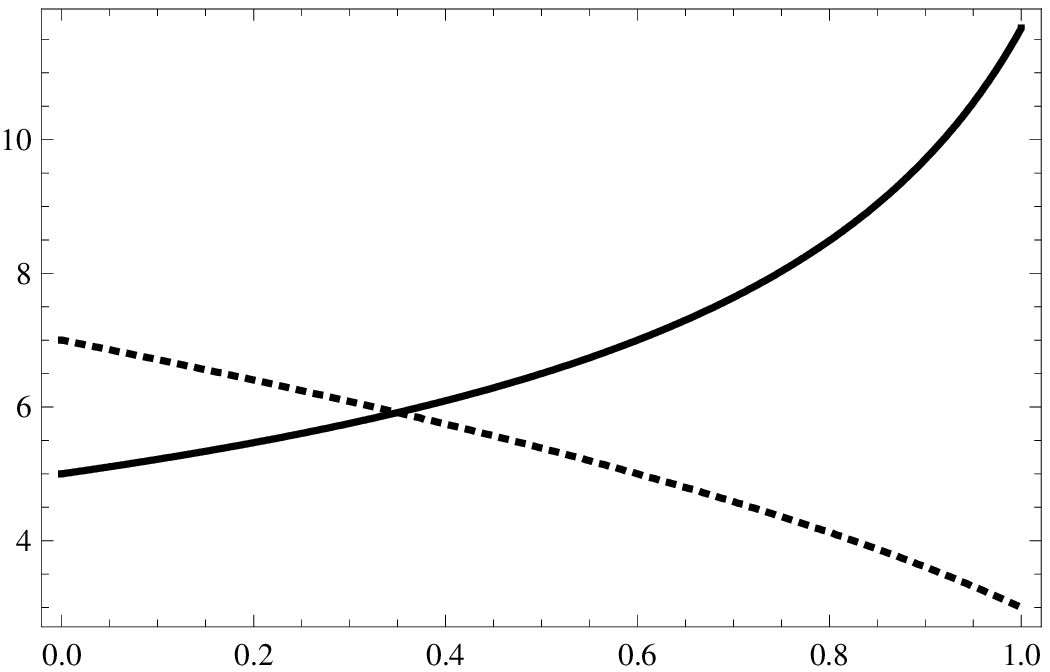} 
\label{warpfigsp1}
}
	\caption{Bach flow : For $\alpha=0\text{(left )}$ and $\beta=0\text{(right)}$ respectively}
	\label{warpfigsp}
	\end{figure}
\subsection{Comparison with Ricci Flow}
We can easily compare the above results with those for un-normalized 
and normalised Ricci 
flows. For our asymmetrically warped metric, the un--normalised and normalised Ricci flow 
equations are, respectively,
\ba\label{urfeqn}
\4{dA^2}{dt}=\pm2\4{A^{2}}{C^{2}}k_{1}\lb( k_{1} + 2 k_{2}\rb) 
\hspace{0.1in};\hspace{0.1in}
\4{dB^2}{dt}=\pm2\4{B^{2}}{C^{2}}k_{2}\lb( k_{1} + 2 k_{2}\rb)
\hspace{0.1in};\hspace{0.1in}
\4{dC^2}{dt}=\pm2\lb( k_{1}^{2} + 2 k_{2}^{2}\rb)
\ea 
\ba\label{nrfeqn}
\4{dA^2}{dt}=\pm \4{A^{2}}{C^{2}}\lb( k_{1} - k_{2}\rb) \lb( k_{1} + 3 k_{2}\rb) 
\hspace{0.01in};\hspace{0.01in}
\4{dB^2}{dt}=\mp \4{B^{2}}{C^{2}}\lb( k_{1} - k_{2}\rb) \lb( k_{1} +  k_{2}\rb) 
\hspace{0.01in};\hspace{0.01in}
\4{dC^2}{dt}=\pm \lb( k_{1} - k_{2}\rb)^{2}
\ea 

The solutions are straight forward and given below--the first set is 
for un--normalised flows and the next one is for normalised Ricci flow.
\ba
A^{2}(t)=A_{0}^{2}{\lb(C_{0}^{2}\pm2\lb(k_{1}^2+2k_{2}^2\rb)t \rb)}^{ k_{1}(k_{1}+2k_{2})/(k_{1}^2+2k_{2}^2)}\\
B^{2}(t)=B_{0}^{2}{\lb(C_{0}^{2}\pm2\lb(k_{1}^2+2k_{2}^2\rb)t \rb)}^{ k_{2}(k_{1}+2k_{2})/(k_{1}^2+2k_{2}^2)}\\
C^{2}(t)~~~=~~~~~~C_{0}^{2}\pm2\lb(k_{1}^2+2k_{2}^2\rb)t 
\ea
\ba
A^{2}(t)=A_{0}^{2}{\lb(C_{0}^{2}\pm2\lb(k_{1}-k_{2}\rb)^{2}t \rb)}^{ (k_{1}+3 k_{2})/(k_{1}-k_{2})}=A_{0}^{2}{\lb(C_{0}^{2}\pm2\lb(k_{1}-k_{2}\rb)^{2}t \rb)}^{
\frac{\alpha}{\gamma}}\\
B^{2}(t)=
B_{0}^{2}{\lb(C_{0}^{2}\pm2\lb(k_{1}-k_{2}\rb)^{2}t \rb)}^{-(k_1+k_2)/(k_1-k_2)}
=B_{0}^{2}{\lb(C_{0}^{2}\pm2\lb(k_{1}-k_{2}\rb)^{2}t \rb)}^{-\frac{\beta}{\gamma}}\\
C^{2}(t)~~~=~~~~~~C_{0}^{2}\pm\lb(k_{1}-k_{2}\rb)^{2}t 
\ea

In the un-normalized Ricci flow, if $k_1=-2k_2$ then $A^{2}(t)$ and $B^{2}(t)$ have no evolution while 
$C^{2}(t)$ grows linearly (the linear behaviour of $C^2(t)$ is 
a point of difference with the corresponding result for Bach flows). 
For normalized Ricci flow, $k_1=k_2$  implies that the R. H. S. of the
flow equations are all zero and there is no evolution. 
Further, for normalized 
Ricci flow  we note that the linear solution for $C^2(t)$ is similar 
to that for un-normalised flow, modulo a difference in the slope.
However, depending on whether $k_1>k_2$ or $k_1<k_2$, 
the solutions for $A^2(t)$ and $B^2(t)$ have mutually opposite behaviour
for normalised flows, though this is not the case for the unnormalised flow. 
We have plotted some of the abovementioned features, in the next figure, 
for forward Ricci flow. Here we have used the same (as for Bach flow) 
initial conditions, i.e. $(A_0^2,B_0^2,C_0^2)=(2,5,7)$ and $(k_1,k_2)=(8,1)$ 
for both un-normalized and normalized flows. 
The figure on the left  
which is for unnormalized Ricci flow, depicts the increasing nature of all the
scale factors though $B^{2}(t)$ asymptotically reaches a constant value. 
This is an immortal flow with a singularity time $T_s=-0.053$. 
If we consider
normalised flows, the  
immortal behavior($T_s=-0.143$) is retained, though all the scale factors 
are not increasing. $A^{2}(t)$ and $C^{2}(t)$ cross each other at some $t$ 
and diverge subsequently. On the other hand, $B^{2}(t)$ decays  and 
asymptotically reaches a constant value. $B^2(t)$ also crosses $A^{2}(t)$ 
and $C^{2}(t)$(not shown in the picture). The figure on the right shows
the abovementioned behaviour for normalised Ricci flow.

The evolution of $A^2(t)$, $B^2(t)$ and $C^2(t)$ for normalised Ricci
flow turns out to be in powers of $t$ which are the squares of 
those for Bach flow. This is evident in the solutions--$A^2(t)$ goes
as $t^\frac{\alpha}{\gamma}$ for normalised Ricci while it varies as
$t^{\frac{\alpha}{2\gamma}}$ for Bach and so on. This is expected
primarily because of the higher order nature of the Bach tensor.

\begin{figure}[htbp]
	\centering
	\subfigure[$A^{2}(t)(continuous)$, $B^{2}(t)(dashed)$, $C^{2}(t)(dotted)$ v/s $t$ ]
{
\includegraphics[width=0.4\textwidth]{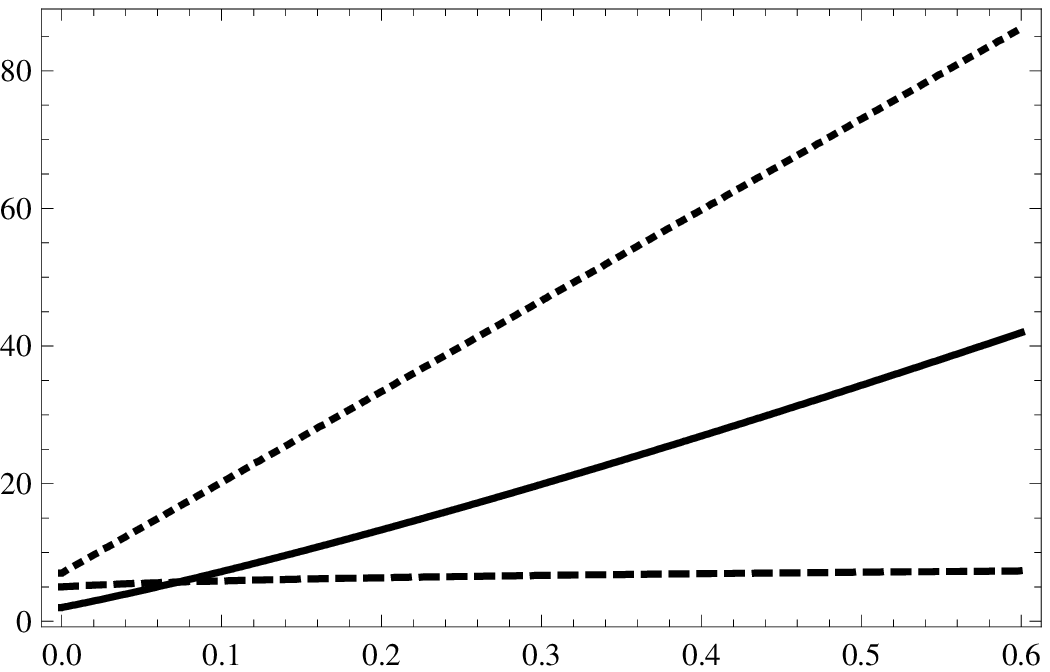} 
\includegraphics[width=0.4\textwidth]{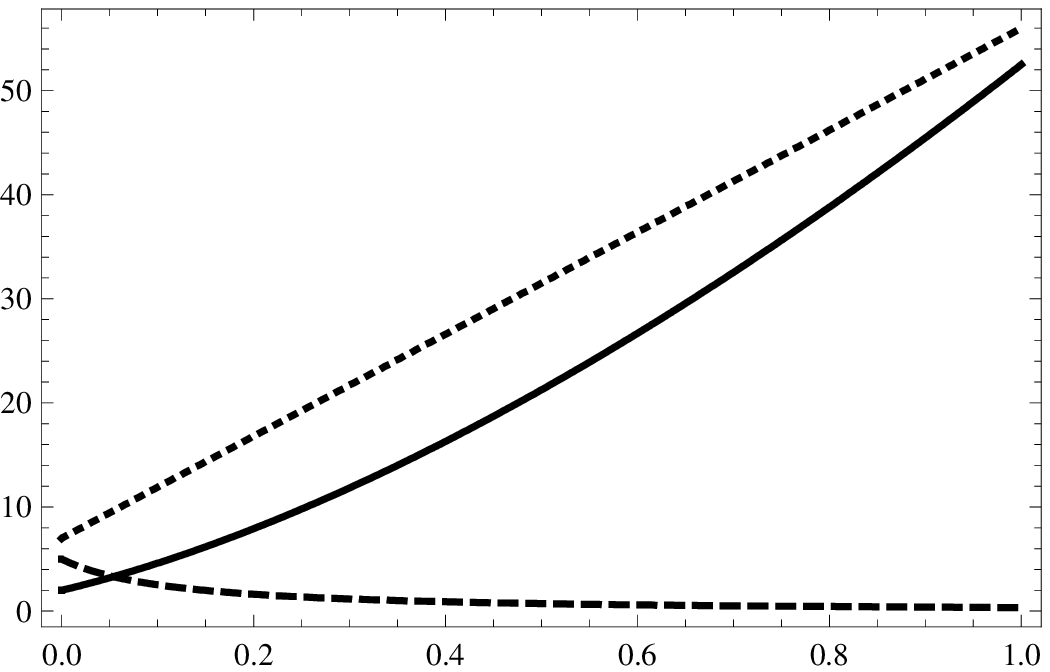} 
\label{figricci1}
}

	\caption{Ricci flow : unnormalized (left) and normalized (right)}
	\label{}
	\end{figure}
\section{Asymmetrically warped products: generalized case}
In this section we briefly re-do our previous analysis for a more general case.
 Let us assume the line element as
\begin{equation}
ds^2=-A^2(t) e^{2 f(\sigma)} d\tau^2 + B^2 (t) e^{2 g(\sigma)}
dx^2 + C^2 (t) e^{2 h(\sigma)}dy^2+ D^2 (t) d\sigma^2
\end{equation}
let us assume further
\begin{equation}
f (\sigma) = k_1 \sigma \hspace{0.2in};\hspace{0.2in} g(\sigma) = k_2 \sigma;\hspace{0.2in} h(\sigma) = k_3 \sigma
\end{equation}
The Bach tensor components for the above mentioned line element are,
\ba
B_{tt}~=~-\4{A^2}{3D^4}\alpha' e^{2 \sigma \kappa _1} \\
B_{xx}~=~-\4{B^2}{3D^4}\beta' e^{2 \sigma \kappa _2}  \\
B_{yy}~=~-\4{C^2}{3D^4}\gamma' e^{2 \sigma \kappa _2}  \\
B_{zz}~=~\4{1}{3D^2}\delta' e^{-2 \sigma  \kappa _3} 
\ea 
where $\alpha', \beta'$ and $\gamma'$ are given as,  
\ba
\alpha'~=~\left(\kappa _1^2+\left(\kappa _2-\kappa _3\right){}^2-2 \kappa _1 \left(\kappa _2+\kappa _3\right)\right) \left(\kappa _1^2-\kappa _2^2-\kappa _2 \kappa _3-\kappa _3^2+\kappa _1 \left(\kappa _2+\kappa _3\right)\right)\\
\beta'~=~ \left(\kappa _1^2-\kappa _2^2-\kappa _2 \kappa _3+\kappa _3^2+\kappa _1 \left(\kappa _3-\kappa _2\right)\right) \left(\kappa _1^2+\left(\kappa _2-\kappa _3\right){}^2-2 \kappa _1 \left(\kappa _2+\kappa _3\right)\right)\\
\gamma'~=~\left(\kappa _1^2+\kappa _1 \kappa _2+\kappa _2^2-\left(\kappa _1+\kappa _2\right) \kappa _3-\kappa _3^2\right) \left(\kappa _1^2+\left(\kappa _2-\kappa _3\right){}^2-2 \kappa _1 \left(\kappa _2+\kappa _3\right)\right)\\
\delta'~=~\left(\kappa _1^2+\left(\kappa _2-\kappa _3\right){}^2-2 \kappa _1 \left(\kappa _2+\kappa _3\right)\right) \left(\kappa _1^2+\kappa _2^2-\kappa _2 \kappa _3+\kappa _3^2-\kappa _1 \left(\kappa _2+\kappa _3\right)\right)
\ea
The tracelessness condition becomes $\alpha'-\beta'-\gamma'+\delta'=0$.\\
If we put $k_{3}=0$ and $B(t)=C(t)$ then the Bach tensor components match with those mentioned in the previous section.
The generic structure of the solutions will not be very different from the 
previous case and we do not discuss them any further here. 

\section{Remarks and conclusions}
We now list our conclusions briefly and make some concluding remarks.

We have solved the Bach flow equations on (2,2) unwarped product manifolds
such as $S^2\times S^2$, $R^2\times S^2$, by making use of the splitting 
of the Bach tensor for such manifolds. The flows conform to the
traceless property of the Bach tensor and are quite different from
Ricci flows on similar manifolds. 

Further, we obtain the fixed point equations for the (2,2) unwarped
product manifolds and have also solved them for restricted cases.
The metric functions and the scalar curvature are found analytically.

Finally, we look at warped product manifolds with a Lorentzian signature.
For a special class of metric functions we reduce the Bach flow
equations to a dynamical system and solve them explicitly.
The evolution patterns of the various flow--parameter dependent
scale factors are obtained and illustrated in detail. Here too the
traceless and higher order characteristics of the Bach tensor
play a role in distinguishing the evolution patterns from those for
Ricci flows. We have compared the Bach flow solutions with those for 
Ricci flow and pointed out the differences. 

It may be mentioned, that the solutions to the flow equations which we have 
obtained in the (2,2) unwarped product manifold cases are
also solutions of $4+1$ dimensional Horava--Lifshitz gravity (for the
restricted choice of the action, as mentioned in the Introduction). 
Another interesting offshoot of our calculations for warped 
manifolds is the Bach flat solution with $k_1=4k_2$ given as
(without the scale factors):
\begin{equation}
ds^2=-e^{8k_2 \sigma} d\tau^2 + e^{2k_2\sigma} \left (dx^2+dy^2\right )
+ d\sigma^2
\end{equation}
This would be a non--singular vacuum solution in conformal gravity 
though we are
not sure whether it is a new solution or a known solution written using 
un--conventional coordinates. 

For future work, it may be worthwhile to find out examples where the higher 
derivative
terms in the Bach tensor play an explicit role in determining 
the flow characteristics.
This is a difficult task considering the fact that evaluating the
Bach tensor is, in itself, quite complicated. Further, one may
consider looking at generalisations of the Bach tensor in other
dimensions \cite{berg} and define newer geometric flows in dimensions
greater than four (note for three dimensions we have the Cotton flow
which has been studied \cite{cotton}). As mentioned before, like other 
geometric flows
the Bach flow is derivable from an action principle (the Weyl--squared
action). It may therefore be possible (and useful) 
to look into an entropy formula and
associated geometric aspects \cite{perelman}. We hope to return to
these issues later.




\begin{references}
\bibitem{bach} R. Bach, \emph{On Weyl's relativity theory and the Weylsian expansion of curved area concepts.}, Math. Ziets. {\bf 9}, 110 (1921). 
\bibitem{berg}J. Bergman, \emph{Conformal Einstein spaces and Bach tensor generalizations in n dimensions}, Ph.d thesis, Matematiska institutionen
Linkopings universitet, Sweden (2004).
\bibitem{ll} L. Landau and E. M. Lifshitz, {\em The Classical Theory of
Fields}, Pergamom Press, UK (1975).
\bibitem{conv} 
Some authors \cite{berg,mann1,mann2} 
use a different sign convention for which the Riemann tensor 
has a sign opposite to what we use here.  The difference in sign
shows up in the contracted Hessian of the Ricci tensor($\nabla^{l}\nabla_{i}R_{lk}=\4{1}{2}\nabla_{i}\nabla_{k}R-R_{il}{R^{l}}_{k}+R_{ijkl}R^{jl}$) and
leads to a different definition of the Bach tensor given as 
$B_{ik} = \nabla^j\nabla^l C_{ijkl} -\frac{1}{2} R^{jl}C_{ijkl}$ 
(note the minus sign).
\bibitem{hlg}I. Bakas, F. Bourliot , D. Lust and M. Petropoulos,\emph{Geometric flows in Horava-Lifshitz gravity}, JHEP {\bf 2010}, No. 4(2009)
\bibitem{ricci} B.~Chow and D.~Knopf, \emph{The Ricci flow: an
introduction},  Mathematical Surveys and Monographs Vol. 110, AMS, 
Providence, 2004.
\bibitem{glik} D. Glickenstein \emph{Lectures on Ricci Flow} Spring 2009, http://math.arizona.edu/~glickenstein/rf/
\bibitem{hamilton} R.S.~ Hamilton, \emph{Three-manifolds with positive Ricci
curvature}, J. Diff. Geom. {\bf 17}, 255 (1982).
\bibitem{friedan} D.~ Friedan, \emph{Nonlinear Models in 2+$\epsilon$
Dimensions}, Annals of Physics {\bf 163}, 318 (1985).
\bibitem{perelman} G. Perelman, \emph{The entropy formula for the Ricci
flow and its geometric applications}, Preprint math.DG/0211159 
\bibitem{bahu} E. Bahuaud and D. Helliwell, \emph{Short time existence for some higher order geometric flows}, Preprint math.10104287v1
\bibitem{mann1}P. D. Mannheim and D. Kazanas, \emph{Newtonian Limit of Conformal Gravity and the Lack of Necessity of the Second Order Poisson Equation },Gen. Rel. and Grav. {\bf 26}, No.4(1994)
\bibitem{mann2}P. D. Mannheim, \emph{Alternatives to Dark Matter and Dark Energy}, Prog. Part. and Nucl. Phys.{\bf 56}, No. 2(2006)
\bibitem{fiedler} B. Fiedler and R. Schimming, \emph{Exact solutions of the 
Bach field equations of general relativity}, Rep. Math. Phys. {\bf 17}, 15 (1980).
\bibitem{yano} K.~Yano, \emph{Differential geometry on complex and almost complex spaces}, Pergamon,. New York, 1965. 
\bibitem{sanjit} S. Das, K. Prabhu, and S. Kar, \emph{Ricci flow of unwarped and warped product manifolds}, Int. Jr. Geom. Meth. Mod. Phys.  
{\bf  7}, 837 (2010)
\bibitem{date} This was pointed out to us by G. Date.
\bibitem{brane} L. Randall and R. Sundrum, \emph{An alternative
to compactification}, Phys.Rev.Lett.{\bf 83},4690 (1999) {\em ibid.} 
\emph{ A large mass hierarchy from a small extra dimension}, Phys.Rev.Lett. {\bf 83}, 3370 (1999)
\bibitem{cotton} A. U. O. Kisisel, O. Sarioglu, B. Tekin, \emph{Cotton flow},
Class.Quant.Grav. {\bf 25}, 165019 (2008).

\end{references}
\end{document}